\documentclass[aps,prb,reprint,floatfix]{revtex4-1}   

\usepackage{amsmath}    
\usepackage{graphicx}   
\usepackage{pstricks}
\usepackage{hyperref}
\usepackage{upgreek}

\newcommand{\unit}[1]{\,\text{#1}}
\newcommand{\outline}[1]{}
\newcommand{\rnd}[1]{\left( #1 \right)}
\newcommand{\sqr}[1]{\left[ #1 \right]}

\begin{document}

\title{Foaming in stout beers}
\author{W. T. Lee} 
\homepage{http://www.ul.ie/wlee}
\email{william.lee@ul.ie}
\affiliation{MACSI, Department of Mathematics and Statistics,
  University of Limerick, Ireland.}  
\author{M. G. Devereux} \affiliation{MACSI, Department of Mathematics
  and Statistics, University of Limerick, Ireland.}


\begin{abstract}
We review the differences between bubble formation in champagne and
other carbonated drinks, and stout beers which contain a mixture of
dissolved nitrogen and carbon dioxide. The presence of dissolved
nitrogen in stout beers gives them a number of properties of interest
to connoisseurs and physicists. These remarkable properties come at a
price: stout beers do not foam spontaneously and special technology,
such as the widgets used in cans, is needed to promote
foaming. Nevertheless the same mechanism, nucleation by gas pockets
trapped in cellulose fibres, responsible for foaming in carbonated
drinks is active in stout beers, but at an impractically slow
rate. This gentle rate of bubble nucleation makes stout beers an
excellent model system for the scientific investigation of the
nucleation of gas bubbles. The equipment needed is very modest,
putting such experiments within reach of undergraduate
laboratories. Finally we consider the suggestion that a widget could
be constructed by coating the inside of a beer can with cellulose
fibres.
\end{abstract}

\maketitle

\section{Introduction}

This paper describes the differences between foaming in drinks
containing dissolved carbon dioxide (sparkling wines, carbonated beers
and soft drinks) of which the most studied example is champagne, and
stout beers. In this work, the term \emph{stout beer} is taken not just to
mean a ``Very dark, full-bodied hopped beer''\cite{Denny2009} but
additionally one which foams due to a combination of dissolved
nitrogen and carbon dioxide. Nitrogen is roughly 50 times less soluble
in water than carbon dioxide and this affects the character
of the bubbles and the foam in the head of the beer. One important
difference is that while champagne and carbonated beers foam
spontaneously, stout beers require special technology to promote
foaming.

It has been shown theoretically and experimentally that stout beers do
foam by the same mechanism as carbonated drinks, but that this occurs
at rate too slow to generate a head in a feasible time. However, the
gentle pace of bubble formation in stout beers make them an ideal
system in which to investigate the physics of nucleation both for
fundamental research and for undergraduate laboratory projects.

\outline{Outline of paper.}  The rest of the article is divided up
into the following sections. Section~\ref{background} summarises the
relevant background physics: Henry's law, Laplace's law and nucleation
barriers.  Section~\ref{champagne} reviews the experimental evidence
that the nucleation sites responsible for bubble formation in
champagne and other carbonated drinks are cellulose fibres and
describes a mathematical model of this nucleation
process. Section~\ref{stout_beers} discusses the unusual properties of
stout beers, most of which be attributed to the low
solubility of the nitrogen gas dissolved in the
beers. Section~\ref{stout_nucleation} describes an extension of the
mathematical model of bubble nucleation in champagne to the case of
two gasses and its application to stout beers. The experimental
observation of this bubbling mechanism is also described. The
implications of these results to future investigations of the science
of nucleation and new technologies for promoting foaming in canned and
bottled stout beers are considered in section~\ref{applications}.
Finally, conclusions are given in section~\ref{conclusions}.

\section{\label{background} Background}

In this section we summarise the background physics used in this
paper, namely Henry's law, which establishes an ``exchange rate''
between the concentration and partial pressure of dissolved gasses;
Laplace's law, which allows us to calculate the overpressure in a gas
pocket trapped within a cellulose fibre; Fick's first law, which
relates the flux of a dissolved gas to its concentration gradient; and
the nucleation barrier to bubble formation in a
supersaturated gas solution. Values of parameters used in the text are
summarised in Table~\ref{parameters}.

\begin{table}[h]
\caption{\label{parameters}Parameters used in the calculations. A
  range of CO$_2$ pressures are used in carbonated beers. Here we
  assume that the total pressure in a carbonated beer is the same as
  in a stout beer.}
\begin{center}
\begin{tabular}{c r@{}l c}
\hline
\hline
Parameter & \multicolumn{2}{c}{Value} & Reference\\
\hline
$r$         &  $6.00\times 10^{-6}$&$\unit{m}$ & \onlinecite{Liger2005b}\\
$\lambda$   & $14.00\times 10^{-6}$&$\unit{m}$ & \onlinecite{Liger2005b}\\
$\gamma$    & $47.00\times 10^{-3}$&$\unit{N}\unit{m}^{-1}$ 
                                              & \onlinecite{Liger2005b}\\
$D_1$       & $1.40\times 10^{-9}$&$ \unit{m}^2 \unit{s}^{-1}$\\ 
$D_2$       & $2.00\times 10^{-9}$&$ \unit{m}^2 \unit{s}^{-1}$\\ 
$H_1$       & $3.4\times 10^{-4}$&$\unit{mol}\unit{m}^{-1}\unit{N}^{-1}$ 
                                                       & \\
$H_2$       & $6.1\times 10^{-6}$&$\unit{mol}\unit{m}^{-1}\unit{N}^{-1}$
                                                       & \\
$T$         & $293$&$\unit{K}$ \\
$P_0$       & $1.00\times 10^{5}$&$ \unit{Pa}$\\ 
\multicolumn{4}{l}{Champagne}\\
$P_1$       & $6.40\times 10^{5}$&$ \unit{Pa}$ & \onlinecite{Zhang2008}\\ 
\multicolumn{4}{l}{Carbonated Beer}\\
$P_1$       & $3.80\times 10^{5}$&$ \unit{Pa}$ & \\ 
\multicolumn{4}{l}{Stout Beer}\\
$P_1$       & $0.80\times 10^{5}$&$ \unit{Pa}$ & \onlinecite{ESGI70}\\ 
$P_2$       & $3.00\times 10^{5}$&$ \unit{Pa}$ & \onlinecite{ESGI70}\\
\hline
\hline
\end{tabular}
\end{center}
\end{table}

\paragraph{Henry's Law} 
states that ``when a gas dissolves the concentration of the (weak)
solution is proportional to the gas pressure.''\cite{LandL5}
Mathematically, $c=HP$, where $c$ is the concentration of the
dissolved gas in solution, $P$ is the partial pressure of the gas in
equilibrium with the solution and $H$ is a constant of
proportionality, the Henry's Law constant. Note that the units of
$c$ may be taken as $\unit{mol}\unit{m}^{-3}$ (this work) or
as $\unit{kg}\unit{m}^{-3}$ and that Henry's law may also be stated as
$P=H^{\prime}c$ (where $H^{\prime}=H^{-1}$) in textbooks. As noted in
the definition, Henry's law only applies to weak (dilute) solutions,
but it is valid for the range of pressures used in this paper.

\paragraph{Laplace's Law}
states that the pressure difference across a curved surface is equal
to $\gamma \left(R_1^{-1} + R_2^{-1}\right)$ where $\gamma$ is the
surface tension of the interface and $R_1$ and $R_2$ are the principal
radii of curvature of the interface.\cite{LandL6} For a spherical
bubble of radius $r$, $R_1=R_2=r$ and the pressure, $P_\text{B}$,
inside the bubble is given by
\begin{equation}
P_{\text{B}}=P_0+\dfrac{2\gamma}{r},
\end{equation}
where $P_0$ is atmospheric pressure. This expression is also valid for
the pressure in a gas pocket trapped within a cellulose fibre, where
$r$ is taken to be the radius of the spherical caps at the ends of the
gas pocket.

\paragraph{Fick's First Law}
describes the flux of a dissolved chemical. Fick's first law is that
the flux $Q$ is proportional to the gradient of the concentration,
$c$, with constant of proportionality $-D$
\begin{equation}
Q=-D\nabla c,
\end{equation}
where, since $D>0$, the minus sign ensures that the flux is directed
from regions of high concentration to regions of low concentration.

\paragraph{Critical Radius} 
Using Henry's law, Laplace's law and Fick's first law we can
understand why there is a nucleation barrier to the growth of
bubbles.\cite{LandL5} Consider a supersaturated gas solution in which
the dissolved gas is at a concentration $c=HP$, where the pressure $P$
of the gas in equilibrium with the solution is greater than
atmospheric pressure $P_0$. Now consider the concentration of that
dissolved gas $c_{\text{B}}$ at the surface of a bubble. We expect
this concentration to be in local equilibrium with the gas in the
bubble
\begin{equation}
\label{bubble_pressure}
c_\text{B}=HP_\text{B}=H\rnd{P_0 +\dfrac{2\gamma}{r}}.
\end{equation}
In order for the bubble to grow, gas must flow from the bulk of the
liquid into the bubble but, by Fick's first law, this will only occur
if $c>c_\text{B}$. Inspection of equation~\ref{bubble_pressure} shows
that this is not guaranteed simply because the solution is
supersaturated, $P>P_0$, but a condition on the size of the bubble
must also be met. The size of the bubble must exceed a critical radius
$r_\text{C}$ given by
\begin{equation}
r_\text{C}=\dfrac{2\gamma}{P-P_0}.
\end{equation}
In other words, for champagne to effervesce and for beer to foam,
bubbles larger than this critical radius must somehow be created.
Bubbles for which $r>r_\text{C}$ are known as post-critical nuclei.
Note that the critical radius was derived using purely mechanical
arguments, its value is a function only of the total pressure of
dissolved gasses in solution independent of, for example, their
solubility.  In the subsequent sections we will discuss how this
nucleation barrier is overcome in practise.

\section{\label{champagne} Bubble nucleation in champagne}

In this paper we take champagne as an example of a drink in which
bubbles form due to dissolved carbon dioxide. This is because
champagne is the most studied of all carbonated beverages, although
the results described below are directly applicable to other
carbonated drinks.

\outline{ Nucleation sites are needed to create bubbles in carbonated
  drinks.}  It has been known for some time that the spontaneous
nucleation of bubbles in carbonated drinks is strongly
inhibited.\cite{Walker1981} As was discussed in the previous section,
only bubbles above a critical radius will grow, and for gas pressures
of less than one hundred atmospheres, the rate of spontaneous
production of these post-critical bubbles either in bulk liquid or on
 surfaces in contact with the bulk liquid (suspended
particles or the container walls) is practically zero.\cite{Jones1999}
Since the pressure of dissolved gasses in champagne is a little over
six atmospheres this mechanism can be ruled out. Nevertheless,
champagne and other carbonated drinks do produce bubbles so an
alternative mechanism must be in play. This is now known to be the
nucleation of bubbles by gas pockets. A gas pocket of post-critical
size will grow due to diffusion of carbon dioxide from solution into
the pocket. When the pocket exceeds some limiting size a bubble
detaches, leaving behind the original, postcritical gas pocket. The
process of growth and detachment repeats resulting in trains of
bubbles rising to the surface.  This is known as type IV
nucleation.\cite{Jones1999}

\begin{figure}[!h]
\begin{center}
\includegraphics{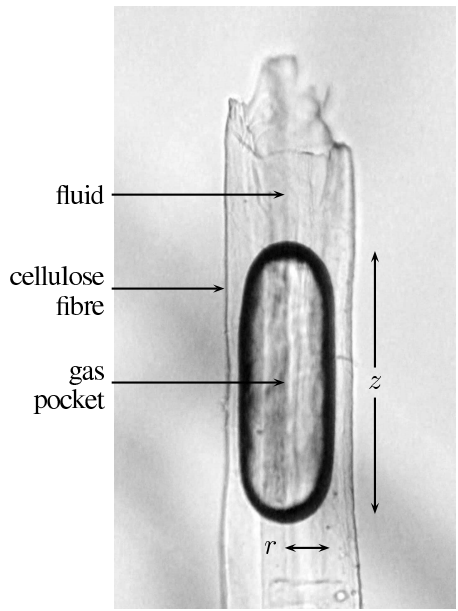}
\caption{\label{geometry} Geometry of a gas pocket trapped in a
  cellulose fibre. The fibre is hollow and contains both a trapped gas
  pocket and fluid. }
\end{center}
\end{figure}

\outline{The surface of glasses are insufficiently rough to create
  bubbles.}  \outline{Observing nucleation sites under a microscope
  reveals cellulose fibres.}  Until relatively recently it was thought
that these gas pockets were held in imperfections on the surfaces of
glasses.  However, examination of the surface topography of glasses
shows that the lengthscales of such imperfections are too small for
gas pockets trapped within them to exceed the critical nucleus size.
Microscopic observation of the sites on champagne glasses from which
trains of bubbles emerge reveal that the gas pockets are trapped
within cellulose fibres as shown in
Figure~\ref{geometry}.\cite{Liger2002} The cellulose fibres will have
either fallen into the glass from suspension in the air or been shed
by the cloth used to wipe the glass dry.

Note that imperfections of a glass surface can nucleate bubbles:
indeed champagne glasses with artificial cavities designed to promote
nucleation are available.\cite{Liger2009b} But in a glass lacking
these artificial cavities, cellulose fibres are more numerous and more
efficient nucleators of bubbles. No bubbles will form in glass with a
smooth surface that has been cleaned with chromic acid solution (which
destroys cellulose and other organic materials).\cite{Taste_of_wine}

\begin{figure}[!h]
\begin{center}
\includegraphics[]{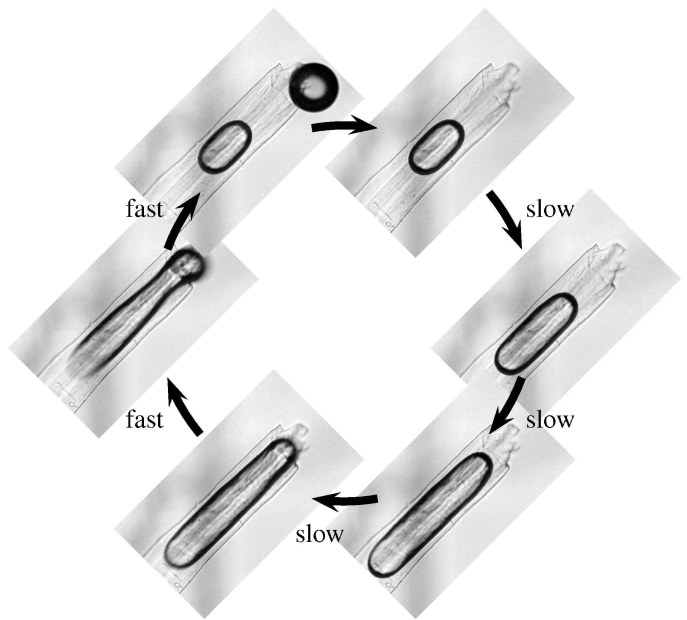}
\caption{\label{nucleation_cycle} Nucleation of a bubble by a gas
  pocket trapped in a cellulose fibre. Although the images taken here
  are from stout beer, the mechanism is exactly the same in carbonated
  liquids. The process of bubble formation has two stages: the slow
  growth of the gas pocket and the fast detachment of a bubble.}
\end{center}
\end{figure}

\outline{The fibre mechanism of bubble nucleation fast and slow
  parts.}  The mechanism of bubble formation by a gas pocket trapped
in a cellulose fibre is illustrated in
Figure~\ref{nucleation_cycle}. A cylindrical gas pocket trapped in a
cellulose fibre grows by diffusion of dissolved gasses into the
pocket. As soon as the pocket reaches the mouth of the fibre it
becomes unstable and a bubble breaks off. There are two timescales to
the process: the timescale of growth of the gas pocket which is
relatively slow and the timescale of detachment of the bubble, which
is fast. It is difficult to say exactly how fast the detachment
process is.  Movies of nucleation events have so far failed to resolve
it. The theory presented below\cite{Liger2005b} predicts the
exponential growth of the gas pocket, and the time constant of that
exponential growth is taken as an estimate of the bubbling time.

\begin{figure}[!h]
\begin{center}
\includegraphics[]{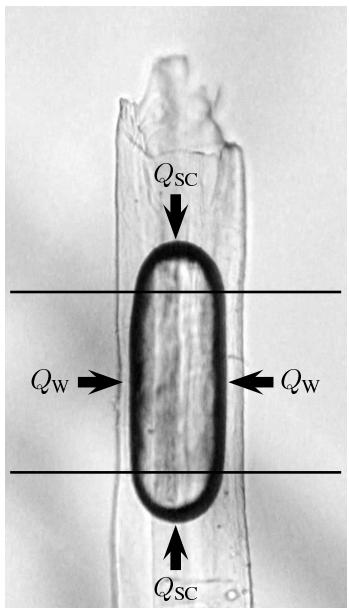}
\caption{\label{flux} The rate of growth of the gas pocket can be
  calculated in terms of $Q_\text{SC}$, the flux of carbon dioxide
  molecules through the spherical caps at the ends of the gas pocket
  and $Q_\text{W}$, the flux through the walls of the cellulose fibre
  into the gas pocket.}
\end{center}
\end{figure}

\begin{figure}[!h]
\begin{center}
\includegraphics[]{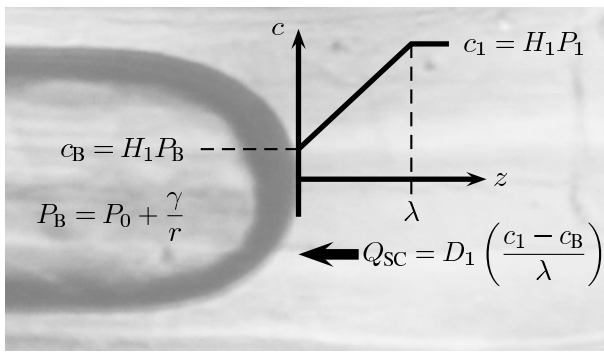}
\caption{\label{diffusion_fig} The diffusion fluxes $Q_\text{SC}$
  (shown) and $Q_\text{W}$ are estimated by introducing a diffusion
  length $\lambda$ over which the carbon dioxide concentration $c$
  varies from the concentration in equilibrium with the gas pocket
  $c_\text{B}$ to the concentration in the bulk liquid
  $c_1$.\cite{Liger2005b}}
\end{center}
\end{figure}

\outline{A mathematical model of fibre nucleation, values typical of
  beers and champagnes.}  A mathematical model of bubble nucleation by
a cellulose fibre can be developed as follows.\cite{Liger2005b} The
rate of change of the amount $N_1$ (in $\unit{mol}$) of carbon dioxide
within the gas pocket can be written in terms of fluxes of carbon
dioxide through the walls of the fibre, $Q_\text{W}$, and through the
spherical caps at the ends of the gas pocket, $Q_\text{SC}$, as shown
in Figure~\ref{flux},
\begin{equation}
\label{ndot}
\dfrac{\text{d} N_1}{\text{d} t}
= 4\pi r^2 Q_\text{SC}
 +2\pi rz Q_\text{W}.
\end{equation}
Two modelling assumptions are used to estimate $Q_\text{SC}$ and
$Q_\text{W}$. The first is illustrated in Figure~\ref{diffusion_fig},
namely that there is a diffusion lengthscale of $\lambda$ over which
the concentration of dissolved carbon dioxide changes from the
concentration in equilibrium with the gas pocket $c_\text{B}$ to the
bulk concentration $c_1$.\cite{Liger2002} This allows us to estimate
\begin{equation}
Q_\text{SC}=D_1 \dfrac{\rnd{c_1-c_\text{B}}}{\lambda}
= \dfrac{D_1H_1}{\lambda}\rnd{P_1-P_\text{B}}.
\end{equation}
The diffusion length $\lambda$ cannot be measured independently so in
practise its value is calculated by working backwards from measured
nucleation rates.\cite{Liger2005b}

The second modelling assumption is used to estimate the flux of carbon
dioxide through the walls of the fibre. The porous walls of the fibre
allow diffusion of carbon dioxide but at a reduced
rate.\cite{Liger2004} This can be modelled by using the same diffusion
length as before and replacing $D_1$ by $D_{1\perp}\approx 0.2
D_1$.\cite{Liger2005b} This allows $Q_\text{W}$ to be estimated as
\begin{equation}
Q_\text{W}=D_{1\perp} \dfrac{\rnd{c_1-c_\text{B}}}{\lambda}
=\dfrac{D_{1\perp}H_1}{\lambda}\rnd{P_1-P_\text{B}}.
\end{equation}
Substituting the above expressions for $Q_\text{SC}$ and $Q_\text{W}$
into equation~\ref{ndot} and using the ideal gas equation,
\begin{equation}
P_\text{B}  \pi r^2 z  = N_1 RT,
\end{equation}
to rewrite $N_1$ in terms of $z$, the length of the gas pocket, gives
\begin{equation}
\dfrac{\text{d} z}{\text{d} t}
=  \rnd{ \dfrac{2D_1}{D_{1\perp}}r + z }            
   \dfrac{2RTH_1D_{1\perp}}{r \lambda}\rnd{\dfrac{P_1}{P_\text{B}}-1}
\end{equation}
Integrating this equation with initial condition $z=z_0$
\begin{gather}
z=\rnd{z_0+\dfrac{2D_{1}}{D_{1\perp}} r }\exp\rnd{\dfrac{t}{\tau}} 
       - \dfrac{2D_{1}}{D_{1\perp}} r\\
\tau= \dfrac{r \lambda P_\text{B}}{2RTH_1D_{1\perp}\rnd{P_1-P_\text{B}}} 
\end{gather}
where $\tau$, the timescale of exponential growth of gas pocket, is
also a good estimate of the time to create a bubble.\cite{Liger2005b}
Substituting in values for champagne gives $\tau=0.04\unit{s}$, and
for carbonated beers $\tau=0.08\unit{s}$.

\section{\label{stout_beers} Bubbles in stout beers}

\outline{Chemical differences between stout beers and carbonated
  drinks.}  There are a number of advantages to using a gas mixture
including a large partial pressure of nitrogen in beers. Replacing
carbon dioxide with nitrogen changes the taste of the beer and the
texture of the head.\cite{Denny2009} Unlike carbon dioxide, nitrogen
is not acidic in solution, giving stout beers a smoother, less bitter
taste. The lower solubility of nitrogen results in smaller bubbles:
these give the head of a stout a creamy mouthfeel. The low
solubility of nitrogen ensures that the head of a stout beer lasts
much longer that of a carbonated beer.\cite{Bamforth2004}

\outline{Waves of falling bubbles} The small bubble size is
responsible for the famous sinking bubbles in stout beers. Drag forces
($\propto r^2$) dominate buoyancy forces ($\propto r^3$) when bubbles
are small and so bubbles are dragged downwards at the sides of the
glass by the circulation of the beer.\cite{Zhang2008} A uniform
downwards flow of bubbles is unstable and so the bubbles cascade down
in waves mathematically equivalent to the roll waves that form in
rapidly flowing shallow water.\cite{Robinson2008}

\outline{Problem also hard to form bubbles. Special equipment needed.}
All the advantages of adding nitrogen come at a price: stout beers do
not foam as readily as carbonated beers. The act of poring a
carbonated beer is enough to generate a head. This is not the case for
stout beers. Special technology is needed to promote foaming in stout
beers. In draught form the beer is forced at high pressure through a
narrow aperture. In cans widgets are used. A widget is a hollow ball
filled with gas at the same pressure as the can. When the can is
opened and the headspace depressurises, the widget also depressurises,
but can only do so through a small submerged nozzle. The widget
releases a turbulent gas jet into the canned beer. The gas
jet rapidly breaks up into the hundred million or so bubble nuclei
needed to form the head. The stirring of the beer generated firstly by
the turbulent jet and secondly by pouring the beer into the glass
ensures that the bubble nuclei mix with the beer, scavenging all the
dissolved nitrogen and carbon dioxide gasses from solution. 

\section{\label{stout_nucleation} Nucleation of bubbles in stout beers}

\begin{figure}
\begin{center}
\includegraphics{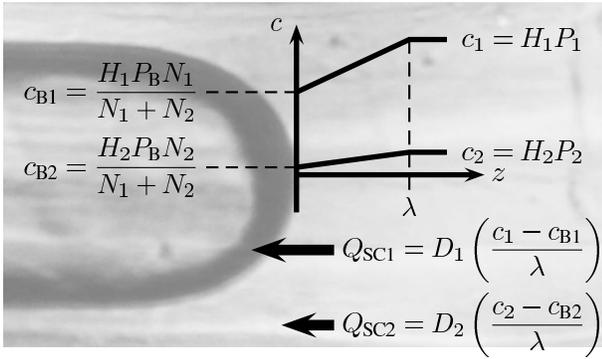}
\end{center}
\caption{\label{second_diffusion_fig}In a stout beer the rate of
  growth of a gas pocket in a cellulose fibre depends on the flux of
  carbon dioxide and nitrogen into the gas pocket. To calculate the
  flux through the spherical cap we assume as before that there is a
  diffusion length $\lambda$ over which the concentrations change from
  their bulk values to those in equilibrium with the bubble. }
\end{figure}

\outline{Interesting to ask why special technology is needed.}  It is
interesting to ask why this special technology is needed. The fibre
nucleation model developed for carbonated liquids can be adapted for
stout beers and used to investigate why the cellulose fibre nucleation
mechanism is insufficient to generate a head.\cite{Lee2011} The gas
pocket in a fibre immersed in stout beer will be at the same total
pressure, $P_\text{B}$, as before, since this only depends on surface
tension and geometry. If the amount of carbon dioxide in the gas
pocket is $N_1$ and the amount of nitrogen is $N_2$ then the partial
pressures of carbon dioxide and nitrogen in the bubble will be given
by
\begin{equation}
P_{\text{B}1}=\dfrac{N_1}{N_1+N_2}P_\text{B}
\qquad
P_{\text{B}2}=\dfrac{N_2}{N_1+N_2}P_\text{B},
\end{equation}
corresponding to surface concentrations of 
\begin{equation}
c_{\text{B}1}=\dfrac{N_1H_1P_\text{B}}{N_1+N_2}\qquad
c_{\text{B}2}=\dfrac{N_2H_2P_\text{B}}{N_1+N_2}.
\end{equation}
Equation~\ref{ndot} becomes two equations for the rate of change of
the amount of carbon dioxide ($N_1$) and nitrogen ($N_2$) in the
gas pocket:
\begin{align}
\label{ndot_one}
\dfrac{\text{d}N_1}{\text{d}t}&=4\pi r^2Q_{\text{SC}1}+2\pi rzQ_{\text{W}1},\\
\label{ndot_two}
\dfrac{\text{d}N_2}{\text{d}t}&=4\pi r^2Q_{\text{SC}2}+2\pi rzQ_{\text{W}2},
\end{align}
where $Q_{\text{SC}1}$ and $Q_{\text{SC}2}$ are the fluxes of carbon
dioxide and nitrogen through the spherical cap of the gas pocket 
as illustrated in Figure~\ref{second_diffusion_fig}
\begin{align}
Q_{\text{SC}1}&=\dfrac{D_1}{\lambda}\rnd{c_1-c_{\text{B}1}}
            =\dfrac{H_1D_1}{\lambda}\rnd{P_1-\dfrac{P_\text{B}N_1}{N_1+N_2}}\\
Q_{\text{SC}2}&=\dfrac{D_2}{\lambda}\rnd{c_2-c_{\text{B}2}}
            =\dfrac{H_2D_2}{\lambda}\rnd{P_2-\dfrac{P_\text{B}N_2}{N_1+N_2}}
\end{align}
The diffusion length $\lambda$ is assumed to be the same for carbon
dioxide and nitrogen.

The fluxes through the the wall of the fibre are given by
\begin{align}
\nonumber
Q_{\text{W}1}&=\dfrac{D_{1\perp}}{\lambda}\rnd{c_1-c_{\text{B}1}}\\
    &=\dfrac{H_1D_{1\perp}}{\lambda}\rnd{P_1-\dfrac{P_\text{B}N_1}{N_1+N_2}},\\
\nonumber
Q_{\text{W}2}&=\dfrac{D_{2\perp}}{\lambda}\rnd{c_2-c_{\text{B}2}}\\
    &=\dfrac{H_2D_{2\perp}}{\lambda}\rnd{P_2-\dfrac{P_\text{B}N_2}{N_1+N_2}}.
\end{align}
We assume that $D_{2\perp}/D_2=D_{1\perp}/D_1=0.2$.

Substituting for the fluxes into equation~\ref{ndot_one}
and~\ref{ndot_two} gives
\begin{multline}
\dfrac{\text{d}N_1}{\text{d}t}
=\dfrac{H_1}{\lambda}
  \rnd{4\pi r^2D_1+2\pi rzD_{1\perp}} \\
  \rnd{P_1-\dfrac{P_\text{B}N_1}{N_1+N_2}},
\end{multline}
\begin{multline}
\dfrac{\text{d}N_2}{\text{d}t}
=\dfrac{H_2}{\lambda}
  \rnd{4\pi r^2D_2+2\pi rzD_{2\perp}} \\
  \rnd{P_2-\dfrac{P_\text{B}N_2}{N_1+N_2}}.
\end{multline}
The ideal gas equation 
\begin{equation}
P_\text{B}\pi r^2 z=\rnd{N_1+N_2}RT
\end{equation}
is used to eliminate $z$ leaving
\begin{multline}
\label{dimensional_one}
\dfrac{\text{d}N_1}{\text{d}t}
=\dfrac{H_1}{\lambda}
  \sqr{4\pi r^2D_1+\dfrac{2D_{1\perp}RT}{rP_{\text{B}}}\rnd{N_1+N_2}}\\
  \rnd{P_1-\dfrac{P_\text{B}N_1}{N_1+N_2}},
\end{multline}
\begin{multline}
\label{dimensional_two}
\dfrac{\text{d}N_2}{\text{d}t}
=\dfrac{H_2}{\lambda}
  \sqr{4\pi r^2D_2+\dfrac{2D_{2\perp}RT}{rP_\text{B}}\rnd{N_1+N_2}}\\
  \rnd{P_2-\dfrac{P_\text{B}N_2}{N_1+N_2}}.
\end{multline}

\subsection{Dimensionless Equations}

To make further progress we rewrite equations~\ref{dimensional_one} and~\ref{dimensional_two} in dimensionless form using scales
\begin{gather}
N_\text{scale}=\dfrac{2D_2P_\text{B}\pi r^3}{D_{2\perp}RT}
               \approx 3.22\times 10^{-13}\unit{mol},\\
t_\text{scale}=\dfrac{rP_\text{B}\lambda}{2D_{2\perp}H_2P_2RT}
               \approx 2.73\unit{s}.
\end{gather}
The dimensionless equations are
\begin{align}
\label{dimensionless_one}
\epsilon \dfrac{\text{d}N_1}{\text{d}t}&=
\rnd{1+N_1+N_2}\rnd{1-\dfrac{\alpha_1 N_1}{N_1+N_2}},\\
\label{dimensionless_two}
\dfrac{\text{d}N_2}{\text{d}t}&=
\rnd{1+N_1+N_2}\rnd{1-\dfrac{\alpha_2 N_2}{N_1+N_2}}.
\end{align}
$N_1$, $N_2$ and $t$ are now dimensionless variables and
\begin{gather}
\epsilon = \dfrac{D_2H_2P_2}{D_1H_1P_1}\approx 0.096,\\
\alpha_1 = \dfrac{P_\text{B}}{P_1}\approx 1.45,\\
\alpha_2 = \dfrac{P_\text{B}}{P_2}\approx 0.39.
\end{gather}

\outline{Solve in the small epsilon limit: maximises insight although
  not quantitatively accurate. } Equations~\ref{dimensionless_one}
and~\ref{dimensionless_two} cannot be solved analytically. They can be
solved using singular perturbation theory\cite{Simmonds1998} in the
asymptotic limit $\epsilon\ll 1$. In this limit
equation~\ref{dimensionless_one} becomes an algebraic equation
\begin{equation}
 0=1-\dfrac{\alpha_1 N_1}{N_1+N_2},
\end{equation}
which can be substituted into equation~\ref{dimensionless_two} to give
\begin{equation}
\dfrac{\text{d} N_2}{\text{d} t}
  = \dfrac{\alpha_1+\alpha_2-\alpha_1\alpha_2}{\alpha_1-1}N_2
   +\dfrac{\alpha_1+\alpha_2-\alpha_1\alpha_2}{\alpha_1}.
\end{equation}
This has the solution
\begin{equation}
N_2=A\exp\rnd{\dfrac{t}{\tau}}
      -\dfrac{\alpha_1+\alpha_2-\alpha_1\alpha_2}{\alpha_1},
\end{equation}
where $A$ is a constant of integration and
\begin{gather} 
\tau=\dfrac{\alpha_1-1}{\alpha_1+\alpha_2-\alpha_1\alpha_2}
     \approx 0.35, \\
\tau t_\text{scale} \approx 0.954\unit{s}.
\end{gather}
Physically the small epsilon limit assumes that equilibration of
carbon dioxide is rapid.

\begin{figure}
\begin{center}
\includegraphics{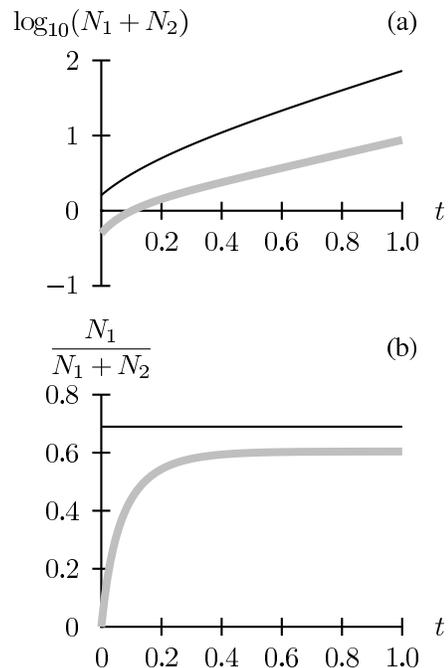}
\caption{\label{plots} Asymptotic (black line) and numerical (grey
  line) solution of equations~\ref{dimensionless_one}
  and~\ref{dimensionless_two}. (a)~Shows the growth of the gas pocket,
  (b)~the mole fraction of carbon dioxide in the bubble. The
  asymptotic solution assumes that the carbon dioxide in the bubble is
  in equilibrium with the solution and overestimates the growth rate
  of the gas pocket.}
\end{center}
\end{figure}

\outline{Compare with numerical solution. Figure~\ref{plots}.}
Equations~\ref{dimensionless_one} and~\ref{dimensionless_two} can be
solved numerically. Figure~\ref{plots} shows the numerical solution
with initial conditions $N_1=0$ and $N_2=0.5$, compared with the
asymptotic solution. Fitting the numerical solution for $5<t<10$ to an
exponential curve gives a dimensionless bubbling timescale of
$\tau=0.47$ and a dimensional timescale of $\tau
t_\text{scale}=1.28\unit{s}$. In other words the cellulose fibre
nucleation mechanism in stout beers is about 15 times slower than in
carbonated beers and about 30 times slower than in champagne. This,
coupled with the fact that the small bubble size in stout beers means
that many more bubbles are needed to make the head, demonstrates why
extra technology is needed to form heads on stout beers.

\subsection{Experimental Observation}

\outline{Experimental observation.}

The model prediction that cellulose fibres can nucleate bubbles in
stout beers can be confirmed
experimentally.\cite{Lee2011,Devereux2011} The first ingredient is
stout beer containing dissolved gasses. If a canned stout is opened
normally, the widget will trigger foaming and the dissolved gasses will
end up in the head of the beer, while the beer itself will no longer
be supersaturated. To avoid this happening the can must be opened
slowly so that the headspace of the can and the widget decompress at
similar rates. This will prevent the widget from creating a large
number of small bubble nuclei and from stirring the liquid.

One procedure for doing this is to put a small amount of bluetack or
other putty-like substance (to create a seal) on the top of the can
and then pierce the can through the bluetack with a pushpin. Be
careful and wear eye protection while doing so. Allow gas to escape
slowly through the hole: leave the pin in place but enlarge the hole
slightly by wobbling it. It should take at least a minute to
decompress the can. The pressure in the head space can be monitored by
squeezing the sides of the can. Once the head space has reached
atmospheric pressure, the pin can be removed. Open the can normally
with the ringpull. Extract supersaturated stout beer from the bottom
of the can with an eyedropper.

If at this point the can is poured into a beer glass then no head will
form and no bubbling will be visible: the beer will be similar in
appearance to flat cola. A strong light directed upwards from the base
of the glass may reveal a few bubble trains nucleated by (invisible)
cellulose fibres.

\begin{figure}[!h]
\begin{center}
\includegraphics[width=6cm]{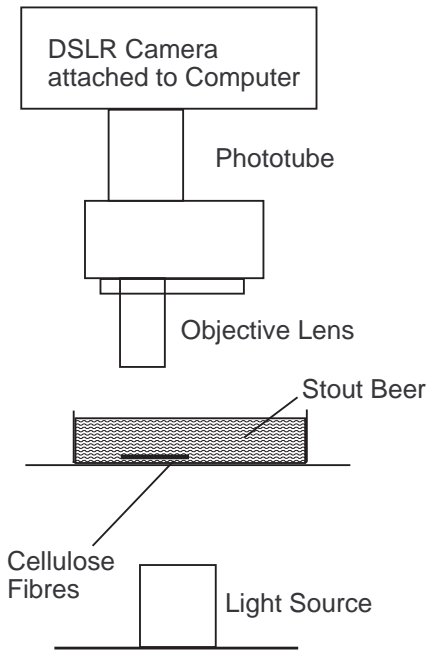}
\caption{\label{experimental_setup} Experimental
  setup.\cite{Devereux2011} Stout beer is poured into a shallow
  container holding cellulose fibres. The nucleation of bubbles is
  viewed through a transmitted light microscope. A digital camera is
  used to record movies of the nucleation events. }
\end{center}
\end{figure}

Figure~\ref{experimental_setup} shows the experimental setup used to
visualise bubble nucleation by fibres.\cite{Devereux2011} Stout beer
is carefully (to minimise agitation) transferred into a shallow
container in which a source of cellulose fibres has been glued to the
base. Laboratory filter paper may be used as the source of the fibres
although coffee filter paper works equally well. The fibres are
observed through a transmitted light microscope. This setup should
be compared with the significantly more sophisticated apparatus needed
to investigate bubble nucleation in champagne.\cite{Liger2006a}

\begin{figure}[!h]
\begin{center}
\includegraphics{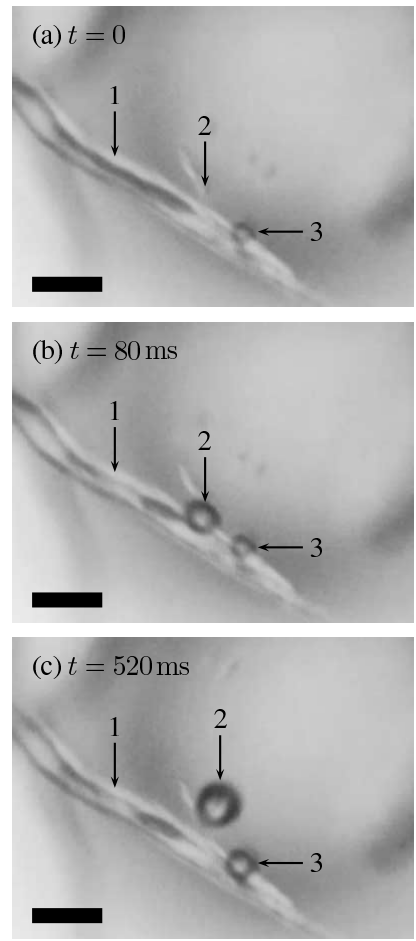}
\caption{\label{nucleation_one} Nucleation of bubbles observed by the
  experimental equipment shown in Figure~\ref{experimental_setup}. The
  figures show (1) a gas pocket trapped in the fibre, (2) a bubble
  created when the air pocket exceeds a critical size, and (3) a
  bubble growing while attached to the fibre. The scale bar is
  $50\,\upmu\text{m}$. (a)~The air pocket (1) has just reached
  critical size. (b)~The air pocket (1) has created a bubble
  (2). (c)~Bubble (2) has visibly detached from the fibre.}
\end{center}
\end{figure}

\begin{figure}[!h]
\begin{center}
\includegraphics{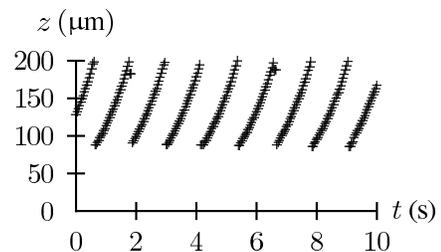}
\caption{\label{observed_length} Plots of length, $z$, of the gas
  pocket shown in Figure~\ref{nucleation_cycle} as a function of
  time. The gas pocket grows slowly, then rapidly contracts as a
  bubble is created.}
\end{center}
\end{figure}

Figure~\ref{nucleation_one} shows several stills from a movie of the
nucleation of bubbles by a fibre. If an unusually large fibre is
viewed, as in Figure~\ref{nucleation_cycle} the growth of the gas
pocket and the formation of the bubble can be seen very clearly. The
length of the gas pocket ($z$) shown in Figure~\ref{nucleation_cycle}
can be measured as a function of time using edge detection. The
results are shown in Figure~\ref{observed_length}. This shows the
steady growth of the gas pocket, as predicted by the model, and the
rapid detachment of the bubble (seen as an almost instantaneous
contraction of the gas pocket) as assumed by the model.

\section{\label{applications} Applications}

\outline{Bubble nucleation in stout beers is easier to observe than
  bubble nucleation in carbonated drinks. }  \outline{Smaller bubbles:
  less agitation of the liquid.No rupture of bubbles at the surface
  coating lens with droplets.}  One interesting result from the
experimental observation of bubble nucleation in stout beers is that
it is much easier to study than in carbonated liquids.  The
experimental setup described above cannot be used to investigate
bubble nucleation in carbonated liquids. The larger bubble size
typical of carbonated liquids leads to agitation of the liquid, moving
the fibres under observation in and out of the plane of focus.  Also
bubbles in some carbonated liquids rupture into a fine spray of
droplets at the surface~\cite{Liger2005} which coat the objective lens
of the microscope, distorting the images. Stout beers therefore are an
ideal model system for fundamental research on the fibre nucleation
mechanism.

\begin{figure}
\begin{center}
\includegraphics{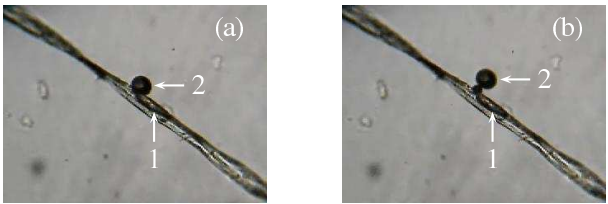}
\end{center}
\caption{\label{usb_microscope} Images of bubble nucleation in stout
  beer by cellulose fibres captured with an entry level usb
  microscope. Despite the unsophisticated equipment, the gas pocket (1)
  and nucleated bubbles (2) can be visualised. }
\end{figure}

The experimental setup is also accessible for undergraduate laboratory
projects. A very minimal set of equipment is needed and good pictures
can be obtained even with an entry level
microscope. Figure~\ref{usb_microscope} shows pictures taken with a
usb microscope sold as a promotional item at a local supermarket. The
images are good enough to observe the gas pockets within individual
fibres as well as the formation of bubbles. Were it not for the
alcoholic nature of the supersaturated solution, these experiments
would also be suitable for school science experiments. Unfortunately,
as noted above, it is much harder to obtain similar results using
carbonated soft drinks.

\outline{Fibres could be used to design a next generation of widgets.}
The fact that cellulose fibres are capable of nucleating bubbles in
stout beers leads to the interesting question of whether an
alternative widget design based on coating the inside of a bottle or
can with fibres would be feasible.\cite{Lee2011} The head of a stout
beer should contain approximately $10^8$ bubbles. In order to produce
these $10^8$ bubbles in the $30\unit{s}$ it takes to pour a stout beer
from the can into a glass, we require $4.3\times10^6$ fibres (each
producing one bubble every $1.28\unit{s}$). If the fibres are arranged
perpendicular to the surface they are attached to, each fibre
occupying an area of $\lambda^2$, the total area needed is then
$8.3\times 10^{-4}\unit{m}^2\approx \rnd{3\unit{cm}}^2$. This rough
estimate suggests such a widget may be feasible.

\section{\label{conclusions} Conclusions}

The bubbles in stout beers have a number of extraordinary properties
which can ultimately be attributed to the low solubility of nitrogen
gas. Stout beers do not foam spontaneously and require special
technology to promote foaming, such as the widget found in cans of
stout beers. However, theory and experiment shows that the same
cellulose fibre nucleation mechanism that causes foaming in carbonated
liquids is also active in stout beers, but at a greatly reduced
rate. The slow nucleation and growth rate of bubbles makes stout beers
an ideal system in which to study nucleation, bringing the study of
nucleation within the reach of undergraduate laboratory equipment. In
the future, coatings of fibres could be applied to the inside of stout
beer cans as a widget replacement.  Future research may focus on
developing a better understanding of the physics behind the empirical
diffusion length $\lambda$ and the bubble detachment process.

\begin{acknowledgments}
We acknowledge support of the Mathematics Applications Consortium for
Science and Industry (\url{http://www.macsi.ul.ie}) funded by the
Science Foundation Ireland Mathematics Initiative Grant 06/MI/005. MGD
acknowledges funding from the Irish Research Council for Science,
Engineering and Technology (IRCSET).
\end{acknowledgments}



\end{document}